\documentclass[fleqn]{article}
\usepackage{gc,amsfonts}

\def\GR{general relativity}
\def\sph{spherically symmetric}
\def\ssph{static, spherically symmetric}

\def\bh{black hole}
\def\bhs{black holes}
\def\asflat{asymptotically flat}

\def\Sch{Schwarzschild}
\def\Tan{Tangherlini}
\def\dS{de Sitter}
\def\adS{anti-de Sitter}

\def\mn{_{\mu\nu}}
\def\MN{^{\mu\nu}}
\def\mN{_\mu^\nu}

\def\og{{\overline g}{}}

\def\cF{{\cal F}}
\def\cK{{\cal K}}
\def\cR{{\cal R}}
\def\ocR{\overline{\cal R}}
\def\tT{{\widetilde T}{}}

\def\M{{\mathbb M}}

\def\R{{\mathbb R}}
\def\S{{\mathbb S}}
\def\T{{\mathbb T}}

\def\ofi{\overline{\phi}}
\def\vfi{\!\vec{\,\varphi}}
\def\Lsc{L_{\rm sc}}

\def\ME {\mbox{$\M_{\rm E}$}}
\def\MED {\mbox{$\M_{\rm E}^D$}}
\def\Mext {\mbox{$\M_{\rm ext}$}}
\def\MJ {\mbox{$\M_{\rm J}$}}
\def\MJD {\mbox{$\M_{\rm J}^D$}}

\heads{K.A. Bronnikov, S.B. Fadeev and A.V. Michtchenko}
{Scalar fields in multidimensional gravity.
                           No-hair and other no-go theorems}

\bls{0.975}
\begin{document}
\thispagestyle{empty}
\twocolumn[
\prepno{gr-qc/0212065}{}

\vspace*{-8mm}

\Title {Scalar fields in multidimensional gravity.\yy
        No-hair and other no-go theorems}

\Aunames{K.A. Bronnikov\auth{a,b,1},
         S.B. Fadeev\auth{a} and A.V. Michtchenko\auth{c,2}}

\Addresses{
\addr a {VNIIMS, 3-1 M. Ulyanovoy St., Moscow 117313, Russia}
\addr b {Institute of Gravitation and Cosmology, PFUR,
        6 Miklukho-Maklaya St., Moscow 117198, Russia}
\addr c {SEPI--ESIME, IPN, Zacatenco, M\'exico, D.F., CP07738, Mexico}
}

\Abstract
     {Global properties of static, spherically symmetric configurations with
     scalar fields of sigma-model type with arbitrary potentials are
     studied in $D$ dimensions, including models where the
     space-time contains multiple internal factor spaces. The latter are
     assumed to be Einstein spaces, not necessarily Ricci-flat, and the
     potential $V$ includes contributions from their curvatures.
     The following results generalize those known in four dimensions:
 (A) a no-hair theorem on the nonexistence, in case $V\geq 0$, of
     \asflat\ \bhs\ with varying scalar fields or moduli fields
     outside the event horizon;
 (B) nonexistence of particlelike solutions in field models with $V\geq 0$;
 (C) nonexistence of wormhole solutions under very general conditions;
 (D) a restriction on possible global causal structures (represented by
     Carter-Penrose diagrams). The list of structures in all models under
     consideration is the same as is known for vacuum with a
     cosmological constant in general relativity: Minkowski (or AdS),
     Schwarzschild, de Sitter and Schwarzschild -- de Sitter, and horizons
     which bound a static region are always simple. The results are
     applicable to various Kaluza-Klein, supergravity and stringy models
     with multiple dilaton and moduli fields.}

] %%%%%%%%%%%%%%%%%%%%%%%%%%%%%%%%%%%%%%%%%%%%%%%%%%%%%%%%%%%%%%%%%%%%
\email 1 {kb@rgs.mccme.ru}
\email 2 {mial@maya.esimez.ipn.mx}

\section {Introduction}    %% S 1

    Extra dimensions have become an inevitable ingredient in numerous
    attempts to build a unification theory including gravity \cite{string}.
    The most popular dimensions are now apparently 10 (superstrings) and 11
    (M-theory), although even higher dimensions are sometimes invoked
    \cite{higher}.  The bosonic sector of such theories generically includes
    scalars (dilatons), and effective scalars (moduli fields) appear
    at dimensional reduction. A diversity of scalar fields are involved in
    other branches of theoretical physics and cosmology:  Goldstone and
    Higgs fields in particle theory, inflatons and scalar dark energy models
    in modern cosmology, etc. It is therefore highly desirable to know the
    possible properties of gravitationally self-bound configurations
    formed by different scalar fields, and of utmost interest are conditions
    for the existence of \bh\ and solitonic, or particlelike solutions.

    The simplest scalar field is massless, minimally coupled to gravity,
    in general, possessing a certain potential.
    Meanwhile, the field equations for self-gravitating scalar fields with
    nontrivial potentials can be explicitly integrated in very few cases,
    even for highly symmetric systems considered in cosmology or for \ssph\
    systems. Therefore, of great value are general inferences or theorems
    on the properties of such systems, which can be obtained without
    entirely solving the field equations.

    For \ssph\ scalar-vacuum configurations in \GR\ (GR),
    described by the action
\bearr                                           \label{act-4}
    S = \int d^4 x \,\sqrt{|g|} \bigl[{\cR}/{\kappa^2} + \Lsc\bigr],
\nnn \cm\cm
    \Lsc = g\MN \varphi_{,\mu} \varphi_{,\nu} - 2V(\varphi),
\ear
    where $\cR$ is the scalar curvature and $\kappa^2$ is the gravitational
    constant, among such theorems are:

\begin{description}    \itemsep 1pt
\item[A.] The no-hair theorem \cite{bek,ad-pear} claiming that
    \asflat\ \bhs\ cannot have nontrivial external scalar fields with
    nonnegative $V(\varphi)$;

\item[B.] The generalized Rosen theorem \cite{brsh91} asserting that
    a particlelike solution (i.e., an \asflat\ solution with a regular
    centre) with positive mass cannot be obtained in case $V\geq 0$;

\item[C.] The nonexistence theorem for regular solutions without a
    centre (e.g., wormholes) \cite{vac1};

\item[D.] The causal structure theorem \cite{vac1}, asserting that
    the list of possible types of global causal structures (and the
    corresponding Carter-Penrose diagrams) for configurations with any
    potentials $V(\varphi)$ and any spatial asymptotics is the same as the
    one for $\varphi = \const$, namely:  Minkowski (or AdS), Schwarzschild,
    de Sitter and Schwarzschild --- de Sitter.
\end{description}

    These results will be referred to as Statements A, B, C, D, respectively.

    A number of exact solutions provide examples of configurations admitted
    by these theorems: \bh\ solutions with a scalar field and $V(\varphi)\geq
    0$ but with non-flat asymptotics \cite{Mann95}, \asflat\ \bh\ and
    particlelike solutions with negative $V(\varphi)$ \cite{vac2}. All this,
    taken together, gives rather a clear picture of what can and what cannot
    be expected from static, minimally coupled scalar fields in \GR.

    There are many generalizations of the action (\ref{act-4}) for which
    it is of interest to know whether or not, or under which additional
    requirements, Statements A--D are valid. In particular:

\begin{enumerate} \itemsep 1pt
\item  %1
    Multidimensional configurations in GR, with $\Lsc$ given by (\ref{act-4})
    and the metric
\beq                                                           \label{ds1}
    ds^2 = \e^{2\gamma} dt^2 - \e^{2\alpha} du^2
                            - \e^{2\beta} d\Omega_{d_0}^2
\eeq
    where $\alpha,\ \beta,\ \gamma$ are functions of the radial coordinate
    $u$ and $d\Omega_{d_0}^2$ is the linear element on the sphere $\S^{d_0}$
    of unit radius.

\item  %2
    More general scalar field Lagrangians in GR, e.g.,
    $\Lsc = F(\varphi,\ I)$ where $ I = g\MN \varphi_{,\mu} \varphi_{,\nu}$
    and $F$ is an arbitrary function of two variables.

\item  %3
    Sets of scalar fields $\vfi = \{\varphi^K\}$, $K= \overline{1,N}$
    of $\sigma$-model type, considered as coordinates in an $N$-dimensional
    target space $\T_{\varphi}$, so that
\beq                                                         \label{L-sigma}
     \Lsc = H_{KL}(\vfi) (\d\varphi^K,\,\d\varphi^L) - 2 V(\vfi)
\eeq
    where the target space metric $H_{KL}$ (usually supposed to be
    positive-definite) and the potential $V$ are functions of $N$ variables
    $\varphi^K$, and we use the notation
\beq
      (\d y,\, \d z) = g\MN \d_\mu y \d_\nu z.                \label{(dd)}
\eeq

\item  %4
    Scalar-tensor theories (STT) of gravity, with the $D$-dimensional action
\bearr\nqq
   S_{\rm STT} = \int d^D x \sqrt{g}                         \label{act-STT}
           [f(\phi) \cR
%%\nnn \inch
           + h(\phi) (\d\phi)^2 -2U(\phi)],
\ear
    where $(d\phi)^2 = (\d\phi,\,\d\phi)$
    and $f,\ h,\ U$ are arbitrary functions of the scalar field $\phi$.

\item  %5
    High-order (curvature-nonlinear) gravity (HOG) theories, e.g., with
    the scalar curvature $\cR$ in (\ref{act-4}) replaced by a function
    $f(\cR)$.

\item  %6
    More general multidimensional configurations, e.g., in product
    manifolds like
\beq                                                         \label{M-prod}
   \M^D = \R_u \times\M_0\times\M_1\times \M_2 \times \cdots \times \M_n
\eeq
    where $\Mext = \R_u \times\M_0 \times\M_1$ is the ``external''
    manifold with the metric (\ref{ds1}), $\R_u \subseteq \R$ is the range of
    the radial coordinate $u$, $\M_1$ is the time axis, $\M_0 = \S^{d_0}$.
    Furthermore, $\M_2, \ldots, \M_n$ are ``internal'' factor spaces of
    arbitrary dimensions $d_i$, $i=2,\ldots, n$, and, according to this
    notation, we also have $\dim \M_0 = d_0$ and $\dim \M_1 = d_1 =1$.
\end{enumerate}

    One can continue the list and/or combine its different items
    to obtain more generalizations.

    As is clear from the previous papers \cite{vac2,vac3,vac4}, some
    extensions are achieved quite easily. Thus, Statements C and D are
    extended to items 1, 2, 3  of the above list in a straightforward manner
    \cite{vac2}.  Extensions of Statements A and B generally require
    additional studies. Statement D proves to be the most universal, in
    particular, it holds \cite{vac3, vac4} in STT and HOG under very general
    conditions. Unlike that, Statement C is violated in STT where wormhole
    solutions are found quite often.

    Let us discuss this point in some detail since it will be relevant in
    what follows.

    A study of STT is effectively conducted with the aid of the well-known
    conformal mapping which generalizes Wagoner's \cite{wagon} 4-dimensional
    transformation
\bear
    g\mn \eql |f(\phi)|^{-2/(D-2)}\og\mn,             \label{g-wag}
\\
    \frac{d\psi}{d\phi} \eql \pm
                \frac{\sqrt{|l(\phi)}|}{f(\phi)}, \label{phi-wag}
\qquad
    l(\phi) \eqdef fh + \frac{D{-}1}{D{-}2}\Bigl(\frac{df}{d\phi}\Bigr)^2,
\ear
    which removes the nonminimal scalar-tensor coupling express\-ed in the
    factor $f(\phi)$ before $\cR$. The action (\ref{act-STT}), originally
    written in the manifold $\MJ[g]$ (the Jordan frame), will now be
    specified in the new manifold $\ME[\og]$ with the new metric $\og\mn$
    (the Einstein frame) and the new scalar field $\psi$:
\bearr \nq\,
    S_{\rm E} = \int  d^D x \sqrt{\og}                   \label{act-E}
    \Bigl\{\sign f \bigl[\ocR
        + (\sign l) (\d\psi)^2\bigr] -2V(\psi)\Bigr\},
\nnn
\ear
    where the determinant $\og$, the scalar curvature $\ocR$ and $(\d\psi)^2$
    are calculated using $\og\mn$, and
\beq
       V(\psi) = |f|^{-D/(D-2)} (\psi)\, U(\phi).             \label{UV}
\eeq
    The action (\ref{act-E}) is similar to (\ref{act-4}), but, in addition
    to arbitrary $D$, contains two sign factors. Let us suppose the usual
    sign of gravitational coupling, $f > 0$. Then theories with
    $l(\phi) < 0$ lead to an anomalous sign of the kinetic term of the
    $\psi$ field in (\ref{act-E}) --- a ``ghost'' scalar field as it is
    sometimes called. Such fields easily violate all the standard energy
    conditions, including the null energy condition, and consequently
    wormholes are quite generic (as was probably first noticed in \Ref{br73}
    in a discussion of \ssph\ solutions to the STT equations with $U=0$).

    Wormholes, however, can even appear in normal STT, with $l(\phi)>0$, due
    to the so-called conformal continuations \cite{vac3}. Namely, it can
    happen that the transformation (\ref{g-wag}) maps a singular surface in
    $\ME$ to a regular surface in $\MJ$ due to a singular behaviour of the
    conformal factor, which compensates the singularity in $\ME$. Then $\MJ$
    can be continued in a regular manner through this surface (the so-called
    {\it conformal continuation} \cite{vac3}), and the global properties of
    $\MJ$ can be considerably richer than those of $\ME$.  \Ref{vac4} has
    established necessary and sufficient conditions for the existence of
    conformal continuations in \ssph\ systems in STT and described the
    properties of conformally continued solutions. It was shown, in
    particular, that a static, traversable wormhole is a generic result of
    conformal continuation.

    In this paper we concentrate our attention on multidimensional
    theories of gravity.

    In \sect 2 we discuss the properties of self-gravitating
    $\sigma$-models with the Lagrangian (\ref{L-sigma}) in space-times
    with the metric (\ref{ds1}). After a brief presentation of the already
    known \cite{vac2} Statements C and D for these systems, we give a proof
    of the no-hair theorem, to a large extent following the ideas of Adler
    and Pearson \cite{ad-pear}. Then, we strengthen Statement B, showing that
    a particlelike solutions of any (even negative) mass cannot be obtained
    in a field model with a nonnegative potential $V(\vfi)$.  This result
    follows from the universal identity (\ref{uni}), valid for all \ssph\
    particlelike configurations, and seems to be new even in four dimensions.

    \sect 3 is devoted to scalar field theories of $\sigma$-model type in
    manifolds of the form (\ref{M-prod}), often obtained in the low-energy
    limit of supergravities, string theories and their generalizations
    \cite{string,higher}. These scalar field theories are reduced to the
    field model studied in \sect 2, with a target space supplemented by
    moduli fields (scale factors of extra dimensions), and
    accordingly the results of \sect 2 are extended to these more general
    theories with certain natural modifications. Moreover, since the theory
    constructed in $\M^D$ (\ref{M-prod}) may be formulated in different
    conformal frames, so that the action takes a form similar to
    (\ref{act-STT}), we briefly discuss the conformal frame dependence of the
    previous results.

    The Appendix contains expressions for some geometric quantities used
    in the previous sections and explicit formulations of the regular
    centre and flat asymptotic conditions for multidimensional space-times.

    Throughout the paper all relevant functions are assumed to be
    sufficiently smooth, unless otherwise explicitly indicated.

\section{Spherically symmetric systems in $D$-dimensional GR
        with a $\sigma$-model source}           %% S2

\subsection{Field equations}                                    %% 2.1

     Consider $D$-dimensional GR with the set of scalar fields
     (\ref{L-sigma}). The Einstein equations can be written in two
     equivalent forms:
\bearr
     G\mN \eqdef \cR\mN - \half \delta\mN \cR = -\kappa^2 T\mN, \label{EE}
\\ \lal \nq
     {\rm or}
\nnn                                                            \label{EE1}
     \cR\mN = - \kappa^2 \tT\mN \eqdef
        - \kappa^2 \biggl(T\mN -\frac{\delta\mN}{D-2}T_\alpha^\alpha\biggr),
\ear
     where $T\mN$ is the stress-energy tensor (SET),
     which for the fields (\ref{L-sigma}) is given by
\beq
     T\mN = \d^\nu\vfi \d_\mu\vfi -\half\delta\mN \Lsc,         \label{Tmn}
\eeq
     or in the ``tilded'' form,
\beq
     \tT\mN = \d^\nu \vfi \d_\mu \vfi -\delta\mN  V(\vfi)      \label{Tmn1}
\eeq
     where two arrows denote a scalar product in the target space
     $\T_{\varphi}$:  $\vec a \vec b = H_{KL} a^K b^L$.

     The \ssph\ metric (\ref{ds1}) is written with an arbitrary radial
     coordinate $u$. Now, it is convenient for our purposes to use the
     coordinate $u=\rho$ corresponding to the gauge condition $\alpha +
     \gamma =0$, so that the metric takes the form
\beq                                                           \label{ds2}
     ds^2 = A(\rho) dt^2 - \frac{du^2}{A(\rho)}
                            - r^2(\rho) d\Omega_{d_0}^2
\eeq
     where we have denoted $r(\rho) = \e^\beta$ and $A(\rho) = \e^{2\gamma}
     \equiv \e^{-2\alpha}$.  This choice is preferable for considering
     Killing horizons, described as zeros of the function $A(\rho)$. The
     reason is that in a close neighbourhood of a horizon the coordinate
     $\rho$ defined in this way varies (up to a positive constant factor)
     like manifestly well-behaved Kruskal-like coordinates used for an
     analytic continuation of the metric \cite{cold}. Thus, using this
     coordinate, which may be called {\it quasiglobal,\/} one can ``cross
     the horizons'' preserving the formally static expression for the
     metric.

     With this choice of the coordinate gauge, the scalar field
     equations and four different combinations of \eqs (\ref{EE1}) can be
     written as follows:
\bear
     \Bigl[A r^{d_0} H_{KL}(\varphi^L)'\Bigr]' \eql
                   r^{d_0} \frac{\d V}{\d\varphi^K};       \label{SE}
\\
    (A'r^{d_0})' \eql - (4/d_0) r^{d_0} \kappa^2 V;        \label{00}
\yy
    d_0 r''/r  \eql  - \kappa^2  (\vfi')^2;                \label{01}
\\yy
         A(r^2)'' - r^2 A'' +  (d_0 \!\lal - 2) r'(2Ar' - A'r)
\nn
                    \eql 2(d_0-1);                 \label{02}
\yy
     d_0(d_0-1)(1-A{r'}^2)\lal - d_0 A'rr'
\nn
              \eql -A r^2(\vfi')^2 + 2r^2 V.   \label{int}
\ear
     \eqs (\ref{00}), (\ref{01}) and (\ref{02}) are the components
     ${t\choose t}$, ${t\choose t} - {\rho\choose \rho}$ and ${t\choose t} -
     {\theta\choose \theta}$, respectively, of (\ref{EE1}), and (\ref{int})
     is the ${\rho\choose\rho}$ component of (\ref{EE}). We have written
     $(N+4)$ equations for $(N+2)$ unknowns $\varphi^K$, $A$ and $r$; but
     there are only two independent equations among (\ref{00})--(\ref{int}),
     in particular, (\ref{int}) is a first integral of the other equations.
     So this set of equations is determined.

\subsection{Global structure theorems}  %% 2.2

     One can directly extend to the present system the reasonings of
     Refs.\,\cite{vac1, vac2} leading to Statements C and D. Let us give,
     for completeness, precise formulations of the corresponding theorems.

     The first theorem concerns the nonexistence of wormholes, horns and
     flux tubes. A {\it wormhole\/} is, by definition, a configuration with
     two asymptotics at which $r(\rho)\to \infty$, hence with $r(\rho)$
     having at least one regular minimum. A {\it flux tube\/} is
     characterized by $r = \const >0$, i.e., it is a static
     $(d_0+1)$-dimensional cylinder. A {\it horn\/} is a configuration that
     tends to a flux tube at one of its asymptotics, which happens if
     $r(\rho)\to\const >0$ at one of the ends of the range of $\rho$. Such
     ``horned particles'' with a flat asymptotic at the other end were
     discussed as possible remnants of black hole evaporation \cite{banks}.

\Theorem{Theorem 1}
   {\eqs (\ref{SE})--(\ref{int}) for $D\geq 4$ and positive-definite $H_{KL}$
     do not admit
       (i) solutions where the function $r(\rho)$ has a regular minimum,
       (ii) solutions describing a horn, and
       (iii) flux-tube solutions with $\varphi^K \neq \const$.
    }

     A proof rests on \eq (\ref{01}), implying $r'' \leq 0$, which actually
     expresses the null energy condition valid for the SET $T\mN$ when the
     matrix $H_{KL}$ is positive-definite. As a result, not only wormholes
     as global entities are impossible but even wormhole throats.

     Another theorem concerns the possible number and order of
     Killing horizons, coinciding with the number and order of zeros of
     $A(\rho)$. A simple (first-order) or any odd-order horizon separates a
     static region, $A>0$ (also called an R region), from a nonstatic
     region, $A < 0$ where (\ref{ds1}) is a homogeneous cosmological metric
     of Kantowski-Sachs type (a T region). A horizon of even order separates
     regions with the same sign of $A(\rho)$.

     The order and disposition of horizons unambiguously determine the
     global causal structure of space-time (up to identification of
     isometric surfaces, if any) \cite{walker}--\cite{strobl}. The following
     theorem severely restricts such possible dispositions.

\Theorem{Theorem 2}
    {Consider solutions to \eqs (\ref{SE})--(\ref{02}) for $D\geq 4$.
    Let there be a static region $a < \rho < b \leq \infty$. Then:
\begin{description}\itemsep -2pt
\item [(i)]
    all horizons are simple;
\item [(ii)]
    no horizons exist at $\rho < a$ and at $\rho > b$.
\end{description} }
\vspace{-1ex}

     A proof of this theorem \cite{vac1, vac2} employs the properties of
     \eq (\ref{02}), which can be rewritten in the form
\beq
     r^4 B'' + (d_0+2)r^3 r'B' = -2(d_0-1)                      \label{02d'}
\eeq
     where $B(\rho) = A/r^2$. At points where $B'=0$, we have $B''<0$,
     therefore $B(\rho)$ cannot have a regular minimum. So, having once
     become negative while moving to the left or to the right along the
     $\rho$ axis, $B(\rho)$ (and hence $A(\rho)$) cannot return to zero or
     positive values.

     By Theorem 2, there can be at most two simple horizons around a
     static region. A second-order horizon separating two nonstatic regions
     can appear, but this horizon is then unique, and the model has no
     static region.

     The possible dispositions of zeros of the function
     $A(\rho)$, and hence the list of possible global causal structures,
     are thus the same as in the case of vacuum with a cosmological
     constant. The latter is a solution to \eqs (\ref{SE})--(\ref{02}) with
     $\varphi^K = \const$, $\kappa^2 V = \Lambda = \const$, and the metric
\bearr
     ds^2 = A(r) dt^2 - \frac{dr^2}{A(r)} - r^2\,d\Omega_{d_0}^2,
                                                              \label{ds2a}
\\ \lal
        A(r) = 1 - \frac{2 m}{(d_0-1) r^{d_0-1}}             \label{A-SdS}
                   - \frac{2\Lambda r^2}{d_0(d_0 + 1)}.
\ear
     This is the multidimensional \Sch-\dS\ (or \Tan-\dS) solution. Its
     special cases correspond to the \Sch\ ($d_0=2$) and \Tan\ \cite{tang}
     ($d_0\geq 2$) solutions%
\footnote
  {The mass $M$ in conventional units, say, grams, is obtained by writing
  $m=GM$ where $G$ is a $(d_0+2)$-dimensional analogue of Newton's
  constant. The coefficient of $m$ is chosen in (\ref{A-Tan}) and
  accordingly in (\ref{A-SdS}) in such a way that at large $r$ in case
  $\Lambda=0$, when the space-time is \asflat, a test particle at rest
  experiences a Newtonian acceleration equal to $-GM/r^{d_0}$.
\label{foot-mass}
  }
     when $\Lambda=0$ and the \dS\ solution in arbitrary dimension when
     $m=0$, called \adS\ (AdS) in case $\Lambda <0$.  For $\Lambda > 0$, if
     $m$ is positive but smaller than the critical value
\beq                                                          \label{mcrit}
    m_{\rm cr} = \frac{d_0-1}{d_0+1}
        \biggl[ \frac{d_0(d_0-1)}{2\Lambda}\biggr]^{(d_0-1)/2},
\eeq
     there are two horizons, the one at smaller $r$ being interpreted as a
     \bh\ horizon and the other as a cosmological horizon. If
     $m=m_{\rm cr}$, the two horizons merge, and there are two homogeneous
     nonstatic regions separated by a double horizon. The solution with
     $m > m_{\rm cr}$ is purely cosmological and has no Killing horizon. In
     cases $m<0$ and/or $\Lambda <0$ there is at most one simple horizon.
     All the corresponding Carter-Penrose diagrams are well known
     (\cite{SdS}, see also \cite{vac3, vac4}) and will not be reproduced
     here.

     In (2+1)-dimensional gravity ($d_0=1$) we have a still shorter list of
     global structures: at most one simple horizon is possible.

     Theorems 1 and 2 are independent of the form of the potential and of
     any assumptions about spatial asymptotics.

\subsection{No-hair theorem}   %% 2.3

     Let us now consider \asflat\ space-times, which means, in terms of the
     metric (\ref{ds2}), that, without loss of generality,
     $r\approx\rho$ and the function $A(\rho) \approx A(r)$ has the \Tan\
     form, i.e., (\ref{A-SdS}) with $\Lambda=0$:
\beq
        A(r) = 1 - \frac{2 m}{(d_0-1)r^{d_0-1}}             \label{A-Tan}
\eeq
     as $\rho\to\infty$. It then follows from the field equations that the
     SET components, and hence the quantities $V$ and $(\vfi)^2$, decay at
     large $\rho\approx r$ quicker than $r^{-(d_0+1)}$.

     Let us now prove the following no-hair theorem, extending to our
     system the theorems known in four dimensions \cite{bek,ad-pear}:

\Theorem{Theorem 3}
    {Given \eqs (\ref{SE})--(\ref{int}) for $D\geq 4$, with a
     positive-definite matrix $H_{KL}(\vfi)$ and $V(\vfi)\geq 0$,
     the only \asflat\ \bh\ solution is characterized by $V\equiv 0$,
     $\vfi = \const$ and the \Tan\ metric (\ref{ds2a}), (\ref{A-Tan})
     in the whole range $h < \rho < \infty$ where $\rho=h$ is the event
     horizon.}

     At the event horizon $\rho=h$ we have by definition $A = A(h) =0$, and
     $A > 0$ at $\rho>h$. By Theorem 2, the horizon should be simple, so
     that $A \sim \rho-h$ as $\rho\to h$. Consider the function
\beq
      \cF_1(\rho) = \frac{r^{d_0}}{r'}[2V - A \vfi'^2].         \label{F1}
\eeq
     One can verify that
\bearr
      \cF_1'(\rho) = \cF_2(\rho) \eqdef                           \label{F2}
        r^{d_0-1} \biggl[2d_0 V + (d_0-1) \frac{\vfi'^2}{r'^2}
                            + A\vfi'^2 \biggr].
\nnn
\ear
     To do so, when calculating $\cF'_1$, one should substitute $\vfi''$ from
     (\ref{SE}), $r''$ from (\ref{01}) and $A'$ from (\ref{int}).
     Let us integrate (\ref{F2}) from $h$ to infinity:
\beq                                                            \label{iden}
      \cF_1(\infty) - \cF_1(h) = \int_{h}^{\infty} \cF_2 (\rho)\, d\rho.
\eeq
     Since $r'(\infty) =1$ and $r'' \leq 0$, we have $r'>1$ in the whole
     range of $\rho$, but $r'(h) < \infty$. Indeed, regularity of the
     horizon implies a finite value of the Kretschmann scalar given
     by (\ref{Kr}), hence finite values of all its constituents (\ref{Kr1}).
     In the present case, the indices $i$ and $k$ (the numbers of factor
     spaces) take the values 0 and 1, and of interest for us is the quantity
     $R_{(3)01} = - \half A'r'$. Since $A'(h) > 0$, its finiteness means
     $r'(h) < \infty$.

     The quantity $\cF_1(h)$ should be finite, since otherwise we would have
     either $V$ or $A\vfi'^2$ infinite, leading to infinite SET components
     (see (\ref{Tmn})) and, via the Einstein equations, to a curvature
     singularity.

     If, however, we admit a nonzero value of $A\vfi'^2$ at $\rho=h$, the
     integral in (\ref{iden}) will diverge at $\rho=h$ due to the second
     term in brackets in (\ref{F2}), and this in turn leads to an infinite
     value of $\cF_1(h)$. Therefore $A \vfi'^2 \to 0$ as $\rho \to h$, and
     we conclude that
\[
     \cF_1(h) = \frac{2r^{d_0}(h)}{r'(h)} V(h) \geq 0.
\]
     On the other hand, $\cF_1(\infty) =0$ due to the asymptotic flatness
     conditions. Thus, in \eq (\ref{iden}) there is a nonpositive quantity
     in the left-hand side and a nonnegative quantity on the right.  The
     only way to satisfy (\ref{iden}) is to put $V\equiv 0$ and $\vfi'
     \equiv 0$ in the whole range $\rho > h$, and the only solution
     for the metric then has the \Tan\ form. $\DAL$

\medskip
     As follows from the scalar field equations (\ref{SE}), the equality
     $V=0$ should take place where $\d V/\d\varphi^K =0$, i.e. at an extremum
     or saddle point of the potential, and it should be obviously a minimum
     for a stable equilibrium.

     It is of interest that one of the key points of the above proof, that
     $A\vfi'^2 =0$ at $\rho=h$, might be obtained from smoothness
     considerations. Indeed, since $A \sim \rho-h$ near $\rho=h$,
     a nonzero value of $A\vfi^2$ means that some of $(\phi^K)'$
     behave as $(\rho-h)^{-1/2}$, violating the $C^1$ requirement for the
     scalar fields. Our proof is ``more economical'' since it only uses the
     requirement of space-time regularity at the horizon.

     One can also note that our no-hair theorem is in a complementarity
     relation with a recent \bh\ uniqueness theorem \cite{roga} (see
     \cite{heus} for a review). In $D$-dimensional \GR\ coupled to the
     $\sigma$-model (\ref{L-sigma}) with $V\equiv 0$, it has been proved
     without assuming spherical symmetry at the outset that ``the only \bh\
     solution with a regular, non-rotating event horizon in an \asflat,
     strictly stationary domain of outer communication is the \Sch-\Tan\
     solution with a constant mapping $\phi$'' \cite{roga}. In contrast to
     that, our Theorem 3 applies to $\sigma$-models with arbitrary
     $V(\vfi)\geq 0$ but selects the \Tan\ solution among \sph\
     configurations.

\subsection{Two expressions for the mass and the properties of particlelike
        solutions}    %% 2.4

     In this subsection we will discuss particlelike solutions, i.e.,
     solutions with a flat asymptotic and a regular centre. We begin with a
     derivation of two general expressions for the active gravitational
     (\Tan) mass $m$ of a $D$-dimensional configuration with the metric
     (\ref{ds1}) and an arbitrary SET compatible with the regular centre
     and asymptotic flatness conditions.

     One expression is easily obtained from the ${t\choose t}$ component of
     \eqs (\ref{EE}) which may be written in the curvature coordinates
     ($u=r$ in the notations of \eqs (\ref{ds1}), (\ref{Rmn})) in
     the following way:
\beq
     \frac{d_0}{(d_0-1)r^{d_0}} \frac{dm}{dr} = \kappa^2 T^t_t,  \label{E00}
\eeq
     where $m(r)$ is the mass function,
\beq
     m(r) \eqdef \frac{d_0-1}{2} r^{d_0-1} (1-\e^{-2\alpha}),  \label{m(r)}
\eeq
     generalizing the well-known 4-dimensional mass function
     $m(r)=\half r(1-\e^{-2\alpha})$. For a system with a regular centre
     ($r=0$), the function $m(r)$, expressed from (\ref{E00}) as
\beq
     m(r) = \frac{d_0-1}{d_0} \kappa^2 \int_{0}^{r} T_t^t r^{d_0}dr,
                                            \label{m-int}
\eeq
     can be interpreted as the mass inside a sphere of radius $r$. If, in
     addition, the space-time is \asflat, this integral converges at large
     $r$ and, being taken from zero to infinity, gives the full \Tan\ mass
     $m=m(\infty)$. The constant $\kappa^2$ is expressed in terms of $d_0$
     and the multidimensional Newtonian constant $G$ (such that $m=GM$, see
     footnote \ref{foot-mass}) if we require the validity of the usual
     expression for mass in terms of density, $M = \int T^t_t\, dv$ ($dv$
     being the element of volume) in the flat space limit.
     One thus obtains
\bear                                                      \label{kappa-G}
     \kappa^2 \eql \frac{d_0}{d_0-1} s(d_0)G,
\nn
     s(d_0) \eql 2\pi^{(d_0+1)/2}\big/\Gamma(\fract{d_0+1}{2}),
\ear
     where $s(d_0)$ is the area of a $d_0$-dimensional sphere of unit radius
     and $\Gamma$ is Euler's gamma function. In case $D=4$ we have, as
     usual, $\kappa^2 = 8\pi G$.

     \eq (\ref{m-int}) for the \Tan\ mass is easily rewritten in terms
     of any radial coordinate, e.g., the quasiglobal $\rho$ coordinate
     used in \eqs (\ref{SE})--(\ref{int}):
\beq                                                          \label{m-Tan}
     m = \frac{d_0-1}{d_0}  \kappa^2
            \int_{\rho_c}^{\infty} T^t_t (\rho) r^{d_0}r'd\rho,
\eeq
     where $\rho_c$ is the value of $\rho$ at the centre.

     On the other hand, one can integrate the ${t\choose t}$ component of
     \eqs(\ref{EE1}), which, in terms of the same $\rho$ coordinate
     (see (\ref{Rmn}) for $R^t_t$), assumes the form
\beq
     \frac{1}{2r^{d_0}} (A'r^{d_0})' = \kappa^2 \tT^t_t.    \label{00-}
\eeq
     For an \asflat\ metric (\ref{ds2}) with a regular center,
     integration of (\ref{00-}) over the whole range of $\rho$ gives
\beq
     m = \frac{\kappa^2}{d_0}                                \label{m-Tol}
        \int_{\rho_c}^{\infty}
            \Bigl [(d_0-1)T^t_t - T^i_i \Bigr ] r^{d_0} d\rho,
\eeq
     where the index $i$ enumerates spatial coordinates.
     This is a multidimensional analogue of Tolman's well-known formula
     \cite{tolman} for the mass of a regular matter distribution in \GR.
     Comparing the expressions (\ref{m-Tan}) and (\ref{m-Tol}), we obtain
     the following {\it universal identity valid for any particlelike
     \ssph\ configuration in $D$-dimensional GR\/}:
\beq                                                        \label{uni}
     \int_{\rho_c}^{\infty}
                \Bigl [(r'-1)(d_0-1)T_t^t + T_i^i \Bigr ] r^{d_0}d\rho =0.
\eeq

     For the $\sigma$-model (\ref{L-sigma}), \eq(\ref{m-Tol}) takes the form
\beq
     m = -\frac{2\kappa^2}{d_0}                                \label{m-phi}
                 \int_{\rho_c}^{\infty} V(\vfi) r^{d_0} d\rho,
\eeq
     leading to a multidimensional version of what has been previously
     called the generalized Rosen theorem \cite{vac2}: {\it a \ssph\
     particlelike solution with positive mass cannot be obtained with scalar
     fields having a nonnegative potential $V$}.

     An even stronger no-go theorem follows from the universal identity
     (\ref{uni}):

\Theorem{Theorem 4}
     {\eqs (\ref{SE})--(\ref{int}) with $D \geq 4$ for the $\sigma$-model
     (\ref{L-sigma}) do not admit any particlelike solution if the matrix
     $H_{KL}$ is positive-definite and $V\geq 0$.}

     In other words, even negative-mass particlelike solutions can only be
     obtained with (at least partly) negative potentials.

     To prove the theorem, it is sufficient to show that the expression in
     brackets in (\ref{uni}) is positive for any nontrivial solution under
     the conditions of the theorem. This expression is
\[
     \half (d_0-1) A(\vfi')^2 + V [2+(d_0-1) r'].
\]
     Its positivity is evident since, as already mentioned, $r'=1$ at the
     flat asymptotic and, due to $r''\leq 0$, we have $r'\geq 1$ in the
     whole range of $\rho$.  $\DAL$

\section{Theories with multiple factor spaces}

\subsection{Reduction}

     Consider a $D$-dimensional \ssph\ space-time $\M^D$ with the structure
     (\ref{M-prod}) and the metric
\bearr \nhq                                                     \label{ds3}
     ds_D^2 = - \e^{2\alpha_0} du^2 - \e^{2\beta^0} d\Omega_{d_0}^2
        + \e^{2\beta^1} dt^2 - \sum_{i=2}^{n}\e^{2\beta^i} ds_i^2,
\nnn
\ear
     where $ds_i^2$ ($i=2, \ldots, n$) are the metrics of Einstein spaces of
     arbitrary dimensions $d_i$ and signatures while $\alpha_0$ and all
     $\beta^i$ are functions of the radial coordinate $u$.

     Consider in $\M^D$ a field theory with the action
\beq
     S = \int d^D x \,\sqrt{|g_D|}                            \label{act-D}
                    \bigl[\cR_D + \Lsc \bigr],
\eeq
     where the scalar field Lagrangian has a $\sigma$-model form similar to
     (\ref{L-sigma}),
\beq
     \Lsc = h_{ab}(\ofi) (\d\phi^a,\ \d\phi^b)-2 V_D(\ofi),  \label{sigma-D}
\eeq
     $\cR_D$ being the $D$-dimensional scalar curvature. The metric $h_{ab}$
     of the $N'$-dimensional target space $\T_{\phi}$ and the potential $V$
     are functions of $\ofi = \{\phi^a\} \in \T_{\phi}$ (we use overbar
s
     for vectors in $\T_{\phi}$ to distinguish them from vectors in
     $\T_{\varphi}$ labelled by arrows). The fields $\phi^a$ themselves are
     assumed to depend on the external space coordinates $x^{\mu}$ ($\mu =
     0, 1, ..., d_0+1$); the notation (\ref{(dd)}) is again used, where the
     metric $g\mn$ is formed by the first three terms in (\ref{ds3}).

     The action (\ref{act-D}) represents in a general form the scalar-vacuum
     sector of diverse supergravities and low-energy limits of string
     and $p$-brane theories \cite{string}. In many papers devoted to
     exact solutions of such low-energy theories (see, e.g.,
     \cite{brane-sol} and references therein), all internal factor spaces
     are assumed to be Ricci-flat, and nonzero potentials $V_D(\ofi)$ are not
     introduced due to technical difficulties of solving the equations.
     Meanwhile, the inclusion of a potential not only generalizes the theory
     making it possible to treat massive and/or nonlinear and interacting
     scalar fields, but is also necessary for describing, e.g.,
     the symmetry breaking and Casimir effects%
\footnote{On the use of effective potentials for describing the Casimir
   effect in compact extra dimensions, see, e.g., \cite{zhuk-cas} and
   references therein.  }.

     Let us perform a dimensional reduction to the external space-time
     \Mext\ with the metric $g\mn$ using the relation (\ref{R-d}). The
     action (\ref{act-D}) is converted to
\bearr                                                       \label{act-J}
        S = \int d^{d_0+2}x \sqrt{|g_{d_0+2}|} \e^{\sigma_2}
     \biggl\{   \cR_{d_0+2}
\nnn \cm
     + \sum_{i=2}^{n} d_i(d_i-1)K_i\e^{-2\beta^i}
     + 2 \nabla^\mu \nabla_\mu \sigma_2
\nnn \cm
     + \sum_{i,k =2}^{n}
            (d_i d_k + d_i\delta_{ik})(\d\beta^i,\d\beta^k) + \Lsc
      \biggr\},
\ear
     where all quantities, including the scalar $\cR_{d_0+2}$, are
     calculated with the aid of $g\mn$, and
\beq
        \sigma_2\eqdef \sum_{d=2}^{n}d_i\beta^i,             \label{sigma2}
\eeq
     so that $\e^{\sigma_2}$ is the volume factor of extra dimensions.

     It is helpful to pass in the action (\ref{act-J}), just as in the
     STT (\ref{act-STT}), from the Jordan-frame metric $g\mn$ in \Mext\ to
     the Einstein-frame metric
\beq
     \og\mn = \e^{2\sigma_2/d_0}g\mn.                       \label{conf}
\eeq
     After this substitution, omitting a total divergence, one obtains the
     action (\ref{act-D}) in terms of $\og\mn$:
\bearr \nhq
     S = \int d^{d_0+2} x \,\sqrt{|\og|}                      \label{act-E}
            \Bigl[\ocR + H_{KL} (\d\varphi^K,\d\varphi^L)
                        -2 V (\vfi)\Bigr].
\nnn
\ear
     Here the set of fields $\{\varphi^K\} = \{\beta^i,\,\phi^a\}$,
     combining the scalar fields from (\ref{sigma-D}) and the moduli fields
     $\beta^i$, is treated as a vector in the extended $N =
     (n{-}1{+}N')$-dimensional target space $\T_{\varphi}$ with the metric
\beq
     (H_{KL}) = \pmatrix {d_i d_k/d_0 + d_i\delta_{ik}  &   0\cr
                  0                      & h_{ab}\cr
             },                                    \label{H-KL}
\eeq
     while the potential $V(\varphi)$ is expressed in terms of $V_D(\ofi)$
     and $\beta^i$:
\beq    \nq                                                     \label{V}
     V(\vfi) = \e^{-2\sigma_2/d_0}\biggl[ V_D(\ofi)
              - \Half \sum_{i=2}^{n} K_i d_i(d_i{-}1)\e^{-2\beta^i}\biggr].
\eeq

     We thus obtain a formulation of the theory coinciding (up to the
     constant $\kappa$ and the particular expression for the potential) with
     that discussed in \sect 2. Therefore all results obtained in \sect
     2 are valid for the metric $\og\mn$ if it assumes the form (\ref{ds1})
     and the quantities $\phi^a$ and $\beta^i$ are functions of the radial 
     coordinate $u$.

\subsection{Extended no-go theorems}

     One can note that the Einstein-frame metric $\og\mn$ in \Mext
     plays an auxiliary role in our multidimensional theory with the action
     (\ref{act-D}). Since the theory is not conformally invariant, the
     physical picture depends on the choice of a conformal frame to be
     regarded as a physical one. This in turn depends on the underlying
     fundamental theory that leads to (\ref{act-D}) in its low-energy limit
     (see \cite{bm01} for a discussion of physical conformal frames and
     further references). We do not specify such a theory, which is possibly
     yet unknown, therefore it seems reasonable to make the simplest choice
     and to consider the properties of the $D$-dimensional metric $g_{MN}$
     given by (\ref{ds3}) as a representative, conditionally physical
     metric. Its ``external'' part $g\mn$ is connected with $\og\mn$ by the
     conformal transformation (\ref{conf}).  Since the action (\ref{act-D})
     corresponds to Einstein gravity in $D$ dimensions, this frame may be
     called the $D$-dimensional Einstein frame, and we will now call $\MED$
     the manifold $\M^D$ endowed with the metric $g_{MN}$.

     The quantities $\alpha_0,\ \beta^0,\ \beta^1$ characterizing $g\mn$ in
     (\ref{ds3}) are connected with $\alpha$, $\beta$, $\gamma$
     corresponding to $\og\mn$ in the form (\ref{ds1}) as follows:
\[  \nhq
     \alpha_0 = \alpha -\sigma_2/d_0,  \quad
     \beta_0  = \beta  -\sigma_2/d_0,  \quad
     \beta_1  = \gamma -\sigma_2/d_0,
\]
     The nonminimal coupling coefficient in the action (\ref{act-J}), being
     connected with the extra-dimension volume factor $\e^{\sigma_2}$, is
     nonnegative by definition, moreover, the solution terminates where
     $\e^{\sigma_2}$ vanishes or blows up. Thus, in contrast to the
     situation in scalar-tensor theories (see the Introduction), conformal
     continuations are here impossible: one cannot cross a surface, if any,
     where $\e^{\sigma_2}$ vanishes. Roughly speaking, due to the
     absence of conformal continuations, the Jordan-frame manifold
     $\Mext [g]$ can be smaller but cannot be larger than $\Mext [\og]$. More
     precisely, the transformation (\ref{conf}) establishes a one-to-one
     correspondence between the two manifolds if $\e^{\sigma_2}$ is regular
     in the whole range $\R_u$ of the radial coordinate in (\ref{ds1}). If
     $\sigma_2 \to \pm \infty$ at an intermediate value of the radial
     coordinate, then the transformation (\ref{conf}) maps $\Mext[g]$ to
     only a part of $\Mext[\og]$.

     As is easily seen from the regularity conditions (\ref{center}) and
     (\ref{flat-as}) presented in the Appendix, the asymptotic flatness of
     the metric $g_{MN}$ in $\MED$ implies an \asflat\ Einstein-frame metric
     $\og\mn$ in \Mext\ and finite limits of the moduli fields
     $\beta^i$, $i\geq 2$, at large $r$. A similar picture is
     observed with the regular centre conditions: a regular centre in $\MED$
     is only possible if there is a regular centre in $\Mext[\og]$ and
     $\beta^i$, $i\geq 2$ behave as is prescribed in (\ref{center}). A
     horizon in $\MED$ always corresponds to a horizon in $\Mext [\og]$.
     (The opposite assertions are not necessarily true, e.g., a regular
     centre in $\Mext[\og]$ may be ``spoiled'' when passing to $g_{MN}$ by
     an improper behaviour of the moduli fields $\beta^i$.)

     So the global properties of $\Mext[\og]$ and $\Mext[g]$ (and
     hence $\MED$) are closely related but not entirely coincide.

     Let us describe some essential features of $\Mext[g]$ and $\MED$,
     associated with Statements A-D in the Introduction.

\medskip\noi
     {\bf A.} The {\bf no-hair theorem} can be formulated for
     $\MED$ as follows:

\Theorem{Theorem 5}
    {Given the action (\ref{act-D}), (\ref{sigma-D}), with $h_{ab}$
     positive-definite and a nonnegative potential (\ref{V}), in the
     space-time $\MED$ with the metric (\ref{ds3}), the only static,
     \asflat\ \bh\ solution to the field equations is characterized
     in the region of outer communication by $\phi^a = \const$, $\beta^i =
     \const$ ($i = \overline{2,n}$), $V(\vfi) \equiv 0$ and the \Tan\ metric
     $g\mn$.
     }

     In other words, the only \asflat\ \bh\ solution is given by the \Tan\
     metric in \Mext, constant scalar fields $\phi^a$ and constant moduli
     fields $\beta^i$ outside the event horizon. Note that in this solution
     the metrics $g\mn$ and $\og\mn$ in $\Mext$ are connected by simple
     scaling with a constant conformal factor since $\sigma_2 = \const$.

     Another feature of interest is that it is the potential (\ref{V})
     that vanishes in the \bh\ solution rather than the original potential
     $V_D(\ofi)$ from \eq (\ref{sigma-D}). If all internal factor spaces are
     Ricci-flat, then both $V_D(\ofi)$ and $V(\vfi)$ are zero in a \bh\
     solution. If not, then the curvatures of the internal factor spaces
     compensate one another or/and the potential $V_D(\ofi)$. The latter,
     if nonzero, is in this case necessarily constant, appearing as a
     cosmological constant in the action (\ref{act-D}).

     Theorem 5 generalizes Theorem 3 from \sect 2 and also the previously
     known property of \bhs\ with the metric (\ref{ds3}) when the internal
     spaces are Ricci-flat and the source is a massless, minimally coupled
     scalar field without a potential \cite{fa91}.

\medskip\noi
     {\bf B. Particlelike solutions:} Theorem 4 is valid in $\MED$
     in the same formulation, but the condition $V\geq 0$ mentioned there
     applies to the potential (\ref{V}) rather than $V_D(\ofi)$ from
     (\ref{sigma-D}).

     One can note that the narrower formulation of the generalized Rosen
     theorem involving the sign of mass could not be so easily extended to
     the metric $g\mn$ since the mass value is, in general, sensitive to
     conformal transformations.

\medskip\noi
     {\bf C. Wormholes} and even wormhole throats are impossible with the
     metric $\og\mn$. The conformal factor $\e^{2\sigma_2/d_0}$ in
     (\ref{conf}) removes the prohibition of throats since for $g\mn$ a
     condition like $r''\leq 0$ (see \eq (\ref{01})) is no longer valid.
     However, a wormhole as a global entity with two flat asymptotics cannot
     appear in $\MJ=\Mext [g\mn]$. Indeed, if we suppose the contrary, then,
     due to the correspondence between flat asymptotics of the two metrics,
     we immediately obtain a wormhole in $\ME = \Mext[\og\mn]$, forbidden by
     Theorem 1.

     Flux-tube solutions with nontrivial scalar and/or moduli fields are
     absent, as before, but horns are not ruled out since the behaviour of
     the metric coefficient $g_{\theta\theta}$ is modified by conformal
     transformations.

     Let us emphasize that all the restrictions mentioned in items A-C are
     invalid if the target space metric $h_{ab}$ is not positive-definite.

\medskip\noi
     {\bf D.} The {\bf global causal structure} of any Jordan frame
     cannot be more complex than that of the Einstein frame even in STT,
     where conformal continuations are allowed \cite{vac4}. The reasoning of
     \cite{vac4} entirely applies to \Mext[g] and hence to $\MED$.
     The list of possible global structures is again the same as that for
     the \Tan-\dS\ metric (\ref{A-SdS}). This restriction does not depend
     (i) on the choice and even sign of scalar field potentials, (ii) on the
     nature of asymptotic conditions and (iii) on the algebraic properties of
     the target space metric. Let us recall that in STT it was also proved
     to be conformal frame independent, regardless of possible conformal
     continuations. It is therefore the most universal property of \sph\
     configurations with scalar fields in various theories of gravity.

\medskip
     A theory in $\M^D$ may, however, be initially formulated in another
     conformal frame, i.e., with a nonminimal coupling factor $f(\ofi)$
     before $\cR_D$ in (\ref{act-D}). Let us designate $M^D$ in this case
     as $\MJD$, a $D$-dimensional Jordan-frame manifold. (An example of
     such a construction is the so-called string metric in string theories
     \cite{string} where $f$ depends on a dilaton field related to string
     coupling.)  Applying a conformal transformation like (\ref{g-wag}), we
     can recover the Einstein-frame action (\ref{act-D}) in $\MED$, then by
     dimensional reduction pass to $\Mext[g]$ and after one more conformal
     mapping (\ref{conf}) arrive at the $(d_0+2)$ Einstein frame
     $\Mext[\og]$. Addition of the first step in this sequence of reductions
     weakens our conclusions to a certain extent. The main point is that we
     cannot {\it a priori\/} require $f(\ofi) > 0$ in the whole range of
     $\vfi$, therefore conformal continuations (CCs) through surfaces where
     $f=0$ are not excluded.

     Meanwhile, the properties of CCs have been studied in \cite{vac4}
     only for a single scalar field in \Mext\ (in the present notation).
     In our more complex case of multiple scalar fields and factor spaces,
     such a continuation through the surface $f(\ofi) =0$ in the
     multidimensional target space $\T_\phi$ can have yet unknown
     properties.

     One can only say for sure that the no-hair and no-wormhole theorems
     fail if CCs are admitted. This follows from the simplest example of CCs
     in the solutions with a conformal scalar field in GR, leading to \bhs\
     \cite{bbm70, bek74} and wormholes \cite{br73,bar-vis99} and known since
     the 70s although the term ``conformal continuation'' was introduced
     only recently \cite{vac3}. Let us also recall that a wormhole was shown
     to be one of the generic structures appearing as a result of CCs in
     scalar-tensor theories \cite{vac4}.

     If we require that the function $f(\ofi)$ should be finite and nonzero
     in the whole range $\R_u$ of the radial coordinate, including its ends,
     then all the above no-go theorems are equally valid in $\MED$ and
     $\MJD$.  One should only bear in mind that the transformation
     (\ref{g-wag}) from $\MED$ to $\MJD$ modifies the potential
     $V_D(\ofi)$ multiplying it by $f^{-D/(D-2)}$, which in turn affects the
     explicit form of the condition $V\geq 0$, essential for Statements A
     and B.

     Statement D on possible horizon dispositions and global causal
     structures will be unaffected if we admit an infinite growth or
     vanishing of $f(\ofi)$ at the extremes of the range $\R_u$. However,
     Statement C will not survive: such a behaviour of $f$ may create a
     wormhole or horn in $\MJD$. A simple example of this kind is a
     ``horned particle'' in the string metric in dilaton gravity of string
     origin, studied by Banks et al. \cite{banks}.

\subsection*{{\large Appendix.}\yy
     Some geometric quantities for $D$-dimensional space-times}

\def\theequation{A\arabic{equation}}
\sequ{0}

     Consider a $D$-dimensional space-time $\M^D$ with the metric
\bearr \nq\,
     ds_D^2 = g_{MN} dx^M dx^N                               \label{ds-D}
     = g\mn dx^{\mu}dx^{\nu} - \sum_{i=2}^{n}\e^{2\beta^i} ds_i^2
%%\nnn
\ear
     where the indices $M,N,\ldots = \overline{0, D-1}$, the indices
     $\mu,\nu$ refer to the external space $\Mext$, while $g\mn$ and
     $\beta^i$ are functions of $x^\mu$.  The internal factor spaces $\M_i$
     ($i=2,...,n$) of arbitrary dimensions $d_i$ and signatures, with
     $x^\mu$-independent metrics $ds^2_i$, are assumed to be Einstein
     spaces, so that the corresponding Ricci tensors can be written as
\beq
     \cR_{m_i}^{n_i} =  K_i (d_i-1)\delta_{m_i}^{n_i},    \label{Ric-i}
\eeq
     where $K_i = +1,\ 0,\ -1$ for spaces of positive, zero and negative
     curvature, respectively. The actual values of their curvatures depend on
     the corresponding scale factors $\e^{2\beta^i}$ in (\ref{ds-D}).

     The scalar curvature of $\M^D$ with the metric (\ref{ds-D}) can be
     expressed in terms of the scalar curvature $\cR_{d_0+2}$ of the
     external subspace $\Mext$ and the functions $\beta^i$:
\bearr                                                           \label{R-d}
     \cR_D = \cR_{d_0+2} + \sum_{i=2}^{n} d_i(d_i-1)\e^{-2\beta^i}
\nnn \nq
     + 2\sum_{i=2}^{n} d_i \nabla^\mu \nabla_\mu \beta^i
     + \sum_{i,k =2}^{n}
            (d_i d_k + d_i\delta_{ik})(\d\beta^i,\d\beta^k)
\ear
     where all quantities, including the Ricci scalar $\cR_{d_0+2}$, are
     calculated with the aid of $g\mn$.

     If the external space-time is \ssph\ with the structure
     $\Mext = \R_u \times \M_0 \times \M_1$
     and the metric (\ref{ds1}) (with the identifications $\alpha=\alpha_0$,
     $\gamma = \beta^1$ and $\beta = \beta^0$), where $\R_u$ is the range of
     the radial coordinate $u$, then the coordinate spheres $M_0 = \S^{d_0}$
     and the time axis $\M_1$ are included in the general scheme, so that
     $K_0 = +1$, $K_1=0$ and $d_1=1$, and the metric can be written as
\bearr
     ds_D^2 =                                               \label{sph-D}
        - \e^{2\alpha_0} du^2 + \sum_{i=0}^{n}\e^{2\beta^i} ds_i^2.
\ear
     In case $n=1$, internal factor spaces are absent, and we return to the
     structure $\M^D = \Mext = \R_u \times \M_0 \times \M_1$.

     Nonzero Ricci tensor components for the \sph\ metric (\ref{sph-D}) are
\bear
     \cR^u_u \eql -\sum_{i=0}^{n}\e^{-2\alpha_0}
        \bigl(\ddot{\beta}^i + \dot{\beta}^i {\mathstrut}^2
                - \dot{\alpha}_0 \dot{\beta}^i\bigr),
\nn
     \cR^{m_i}_{n_i} \eql \delta^{m_i}_{n_i} \biggl\{
                    (d_i-1)K_i \e^{-2\beta^i}
\nnn \quad
       -\e^{-2\alpha_0} \biggl [\ddot{\beta}^i
     + \dot{\beta}^i\biggl(-\dot{\alpha}_0 +\sum_{k=0}^{n}d_k\dot{\beta}^k
            \biggr)\biggr]\biggr\},
\ear
     where the dots denote $d/du$ and the indices $m_i$, $n_i$ belong to
     coordinates from the $i$-th factor space.

     From these general expressions, putting $n=1$, it is easy to obtain
     the Ricci tensor components in the particular gauge $\alpha_0+\beta^1=0$
     (the quasiglobal coordinate $\rho$) for the metric (\ref{ds2}) used in
     \sect 2:
\bear                                                          \label{Rmn}
     \cR_t^t \eql -\frac{1}{2r^{d_0}}(A' r^{d_0})';
\nn
     \cR_\rho^\rho   \eql -\frac{1}{2r^{d_0}}(A' r^{d_0})'
                    -d_0 A \frac{r''}{r};
\nn
     \cR_\theta^\theta
             \eql \frac{d_0-1}{r^2} - A
       \biggl[\frac{r''}{r} + (d_0-1) \frac{r'^2}{r^2}
          + \frac{r'}{r}\,\frac{A'}{A} \biggr],
\ear
     where $A = \e^{2\beta^1}$, $r = \e^{\beta^0}$, $\theta$ is any of the
     angular coordinates parametrizing the sphere $\S^{d_0}$ and the prime
     denotes $d/d\rho$.

     Let us now return to the metric (\ref{sph-D}) and give an
     expression for its Kretschmann scalar (Riemann tensor squared)
     $\cK = \cR_{MNPQ}\cR^{MNPQ}$:
\bearr  \nq                                                     \label{Kr}
     \cK = 4\sum_{i} R_{(1)i}^2
            + 2\sum_{i} d_i(d_i{-}1) R_{(2)i}^2
                + 4\sum_{i\ne k} d_i d_k R_{(3)ik}^2
\nnn
\ear
     where $i, k = \overline{0,n}$ and
\bear
     R_{(1)i}  \eql - \e^{-2\alpha}
     (\ddot\beta^i -\dot{\alpha}\dot{\beta}^i + \dot\beta^i{\mathstrut}^2),
\nn
     R_{(2)i}  \eql K_i \e^{-2\beta^i}
            - \e^{-2\alpha} \dot\beta^i{\mathstrut}^2,
\nn
     R_{(3)ik} \eql - \e^{-2\alpha} \dot{\beta}^i\dot{\beta}^k;  \label{Kr1}
\ear
     as before, the dots denote $d/du$ (recall that $u$ is an arbitrary
     radial coordinate, not to be confused with the particular coordinate
     $\rho$ in \eqs (\ref{Rmn})).

     The expression (\ref{Kr}) is a sum of squares of the Riemann
     tensor components $\cR \mn{}^{\lambda\sigma}$, hence its proper
     behaviour at the centre or any other point guarantees the regularity
     properties of the manifold $\M^D$. Thus, a centre is by definition a
     place where the coordinate spheres are drawn to points, i.e., $r =
     \e^{\beta_0}\to 0$.  The regular centre conditions follow from the
     requirement $\cK < \infty$ at such a value of the coordinate $u$:
\bearr \nq  \,                                               \label{center}
     \beta^i = \beta^i_0 + O(r^2), \quad i = \overline{1,n};
  \quad\
     \e^{-2\alpha}\dot{r}{}^2 = 1 + O(r^2),
\nnn
\ear
     where $\beta^i_0$ are constants. The last condition is the local
     euclidity requirement, providing a correct circumference to radius
     ratio for circles around the centre; it follows from the
     requirement of finiteness of $R_{(2)0}$ at $r=0$.

     The metric (\ref{sph-D}) can be called \asflat\ when its external part
     is \asflat\ and the internal scale factors $\e^{\beta^i}$ tend to
     finite constant values as $r\to\infty$. Since $\beta^i$ behave as
     effective scalar fields in \Mext, the asymptotic flatness of \Mext\
     requires a sufficiently rapid decay of $\dot{\beta}^i$ at infinity.
     We thus require
\bear
     \beta^i = \beta^i_\infty + O(1/r), \quad i = \overline{1,n};
  \quad\
     \e^{-2\alpha}\dot{r}{}^2 \to 1.                       \label{flat-as}
\ear
     where $\beta^i_\infty$ are constants. The last condition in
     (\ref{flat-as}) is the requirement of a correct circumference to radius
     ratio for circles $r=\const$ as $r\to \infty$; it rules out the
     possibility of asymptotics of conical nature. Under these conditions,
     all components of the Riemann tensor manifestly vanish at infinity.

     The conditions (\ref{center}) and (\ref{flat-as}) are written for an
     arbitrary radial coordinate $u$ but may be easily reformulated for its
     particular choice. Thus, for the quasiglobal coordinate $\rho$ one can
     assume $r\approx \rho$ at large $r$, then the last condition in
     (\ref{flat-as}) reads simply $\e^{-2\alpha_0} \equiv  A \to 1$ as
     $\rho\to\infty$. The central value of $\rho$ is $\rho_c$, not
     necessarily zero; as follows from the first condition (\ref{center}),
     $A(\rho_c) = A_c$ is finite, while the last condition yields
     $A(dr/d\rho)^2 = 1 + O\bigl((\rho-\rho_c)^2\bigr)$ near the centre.

\subsection*{Acknowledgements}

We are grateful to Irina Dymnikova and Vitaly Melnikov for helpful
discussions. KB and SF acknowledge partial financial support from the
Russian Foundation for Basic Research and the Russian Ministry of Industry,
Science and Technologies.


\small
\begin{thebibliography}{99}   \itemsep 1pt
\bibitem{string}
        A. Salam and E. Sezgin, eds., ``Supergravities in
        Diverse Dimensions", reprints in 2 vols., World Scientific, 1989;
      \\
        M.B. Green, J.H. Schwarz, and E. Witten, ``Superstring
        Theory'' in 2 vols., Cambridge Univ. Press, 1987;
      \\
        K.S. Stelle, ``Lectures on supergravity p-branes '', hep-th/9701088;
      \\
        M.J. Duff, ``M-theory (the theory formerly known as strings)'',
        hep-th/9608117.

\bibitem{higher}
    N. Khviengia, Z. Khviengia, H. L\"u and C.N. Pope,
        ``Toward a field theory of F-theory'',
    \CQG {15} 759 (1998); hep-th/9703012;
      \\
    I. Bars, \PLB {403} 257 (1997);
      \\
    A.M. Gavrilik, {\it Acta Phys. Acad. Sci. Hung. }{\bf 11}, 35 (2000).

\bibitem{bek}
        J.D. Bekenstein, \PRD {5} 1239 (1972); ibid., 2403;
        ``Black holes: classical properties, thermodynamics, and heuristic
    quantization'', gr-qc/9808028 (review).

\bibitem{ad-pear}
    S. Adler and R.B. Pearson, \PRD {18} 2798 (1978).

\bibitem{brsh91}
        K.A. Bronnikov and G.N. Shikin, ``Self-gravitating  particle
        models with  classical  fields and their stability''.  Series
        ``Itogi Nauki i Tekhniki" (``Results of Science and Engineering''),
        Subseries ``Classical Field Theory and Gravitation Theory", v.\,2,
        p. 4, VINITI, Moscow 1991 (in Russian).

\bibitem{vac1}
        K.A. Bronnikov, \PRD {64} 064013 (2001).

\bibitem{Mann95}
        K.C.K. Chan, J.H. Horne and R.B. Mann, \NPB {447} 441 (1995).

\bibitem{vac2}
        K.A. Bronnikov and G.N. Shikin,
        {\it Grav. \& Cosmol\/} {\bf 8}, 107 (2002); gr-qc/0109027.

\bibitem{vac3}
        K.A. Bronnikov, {\it Acta Phys. Polon. } {\bf B32}, 3571 (2001);
        gr-qc/0110125.

\bibitem{vac4}
        K.A. Bronnikov, \JMP {43} No. 12 (2002), gr-qc/0204001.

\bibitem{wagon}
        R. Wagoner,
%%      ``Scalar-tensor theory and gravitational waves.''
        {\it Phys. Rev. } {\bf D 1}, 3209 (1970).

\bibitem{br73}
        K.A. Bronnikov,
%%      ``Scalar-tensor theory and scalar charge''
        {\it Acta Phys. Polon. } {\bf B4}, 251--273 (1973).

\bibitem{cold}
        K.A. Bronnikov, G. Cl\'ement, C.P. Constantinidis and J.C. Fabris,
        \PLA {243} 121 (1998), gr-qc/9801050;
        \GC {4} 128 (1998), gr-qc/9804064.

\bibitem{banks}
        T. Banks, A. Dabholkar, M.R. Douglas and M. O'Loughlin,
            {\it Phys. Rev. } {\bf D 45}, 3607 (1992);\\
        T. Banks and M. O'Loughlin,
            {\it Phys. Rev. } {\bf D 47}, 540 (1993).

\bibitem{walker}
        M. Walker,
%%      ``Block diagrams and the extension of timelike two-surfaces.''
         {\it J. Math. Phys.\/} {\bf 11}, 8, 2280 (1970).

\bibitem{br79}
        K.A. Bronnikov,
        {\it Izv. Vuzov, Fizika } No. 6, 32 (1979).

\bibitem{katan}
        M.O. Katanaev,
        {\it Nucl. Phys. Proc. Suppl.\/} {\bf 88}, 233--236 (2000),
            gr-qc/9912039;
        {\it Proc. Steklov Inst. Math.\/} {\bf 228}, 158--183,
            gr-qc/9907088.

\bibitem{strobl}
        T. Klosch and T. Strobl, \CQG {13} 2395--2422 (1996);
        {\bf 14}, 1689--1723 (1997).

\bibitem{tang}
    F.R. Tangherlini, {\it Nuovo Cim.\/} {\bf 77}, 636 (1963).

\bibitem{SdS}
        K. Lake and R. Roeder, \PRD {15} 3513 (1977);\\
        M. Katanaev, T. Klosch and W. Kummer.
%%      ``Global properties of warped solutions in general relativity'',
        {\it Ann. Phys. (USA)} {\bf 276}, 191 (1999).

\bibitem{roga}
    M. Rogatko, ``Uniqueness theorem for static black hole solutions
    of $\sigma$-models in higher dimensions'', hep-th/0207187.

\bibitem{heus}
    M. Heusler, ``Black Hole Uniqueness Theorems''.
        Cambridge Univ. Press, 1997;\\
    P.O. Mazur, ``Black Hole Uniqueness Theorems'', hep-th/0101012.

\bibitem{tolman}
    R.C. Tolman, ``Relativity, Thermodynamics and Cosmology'',
    Clarendon Press, Oxford, 1969.

\bibitem{brane-sol}
        K.A. Bronnikov, V.D. Ivashchuk and V.N. Melnikov, gr-qc/9710054;
        {\it Grav. and Cosmol.\/} {\bf 3}, 203 (1997);\\
        K.A. Bronnikov,
        %%``Block-Orthogonal Brane Systems, Black Holes and Wormholes",
        hep-th/9710207;
        {\it Grav. and Cosmol.\/} {\bf 4}, 49 (1998);\\
        V.D. Ivashchuk and V.N. Melnikov, \CQG {18} R87--R152 (2001).

\bibitem{zhuk-cas}
        U. G\"unther and A. Zhuk, \PRD {56} 6391 (1997);\\
        U. G\"unther, S. Kriskiv and A. Zhuk, \GC {4} 1 (1998).

\bibitem{bm01}
        K.A. Bronnikov and V.N. Melnikov,
%%      ``On observational predictions from multidimensional gravity'',
        gr-qc/0103079, \GRG {33} 1549 (2001).

\bibitem{fa91}
        S.B. Fadeev, V.D. Ivashchuk and V.N. Melnikov,
        \PLA {161} 98 (1991).

\bibitem{bbm70}
        N.M. Bocharova, K.A. Bronnikov and V.N. Melnikov,
        {\it Vestn. Mosk. Univ., Fiz. Astron. } No. 6, 706 (1970).

\bibitem{bek74}
        J.D. Bekenstein,
%%      ``Exact solutions of Einstein-conformal scalar equations.''
        {\it Ann. Phys. (USA)\/} {\bf 82}, 535 (1974).

\bibitem{bar-vis99}
    C. Barcel\'o and M. Visser,
    \PLB {466} 127--134 (1999).

\end{thebibliography}
\end{document}